\documentclass[article,aps]{revtex4}  

\usepackage{amsmath}    
\usepackage{graphicx}   
\newcommand{\ket}[1]{\left|#1\right\rangle}

\begin{document}

\title{Quantum computer networks with the orbital angular momentum of light}
\author{Juan Carlos Garcia-Escartin}
 \email{juagar@tel.uva.es}  
\affiliation{Universidad de Valladolid, Dpto. Teor\'ia de la Se\~{n}al e Ing. Telem\'atica, Paseo Bel\'en n$^o$ 15, 47011 Valladolid, Spain}
\author{Pedro Chamorro-Posada}
\affiliation{Universidad de Valladolid, Dpto. Teor\'ia de la Se\~{n}al e Ing. Telem\'atica, Paseo Bel\'en n$^o$ 15, 47011 Valladolid, Spain}
\date{\today}

\begin{abstract}
Inside computer networks, different information processing tasks are necessary to deliver the user data efficiently. This processing can also be done in the quantum domain. We present simple optical quantum networks where the orbital angular momentum of a single photon is used as an ancillary degree of freedom which controls decisions at the network level. Linear optical elements are enough to provide important network primitives like multiplexing and routing. First we show how to build a simple multiplexer and demultiplexer which combine photonic qubits and separate them again at the receiver. We also give two different self-routing networks where the OAM of an input photon is enough to make it find its desired destination.  
\end{abstract}
\maketitle

\section{Quantum computer networks}
Classical communication networks are routinely used to coordinate many users in common computation and communication tasks. The successful implementation of quantum communication protocols, like quantum cryptography \cite{AJP03,SLB11}, has spawned an interest in quantum communication networks. In this Letter, we explore the possibilities of photonic quantum computer networks which work with the orbital angular momentum of light. 

When many computers are connected to a network, there appear new problems which are dealt with at different levels \cite{Sta00}. Two important problems are routing and multiplexing. Routing concerns finding a path from the origin to the destination node. The network must have a mechanism to identify the destination and select the best way to deliver the data. Multiplexing protocols control how the limited network resources are shared among the users. For instance, in fibre optic networks, Wavelength Division Multiple Access, WDMA, schemes are employed to send many data channels through the same optical fibre. The signals are sent at different wavelengths so that they not interfere with each other. 

There are also quantum techniques to combine different quantum channels in the same path. The channels are given separate properties in the time \cite{CYT11}, frequency and wavelength \cite{BBG03,OC06,MRA12}, coherent state amplitude \cite{GC09} or orbital angular momentum \cite{GC08} domains so that they can be later divided again at the receiver.

Additionally, quantum computation permits new routing protocols, like delayed commutation \cite{GC06b}, that can solve certain problems of classical networks.

In order to perform network tasks like routing and multiplexing, the network usually carries two kinds of information: the user data which must be sent from one end to the other and the control data necessary for the correct operation of the network.

We will discuss networks in which the control data is encoded in the orbital angular momentum of single photons. The user data can be encoded in another degree of freedom. In our networks, we use polarization. With this separation, we can work on the network problems at a level independent from the actual transmitted data. 

Our schemes are based on an elementary optical block, the orbital angular momentum beam splitter, OAMBS, which can be built using only mirrors, phase shifters, beamsplitters and Dove prisms \cite{ZM05}. OAMBSs, when combined with computer generated holograms, can be used to provide a simple OAM multiplexing scheme and two kinds of self-routing networks where the data is directed to its destination without advanced processing in the intermediate nodes. The schemes we present are also valid for classical networks. We take the more general quantum case because it also allows new possibilities like routing to a superposition of destinations. 
 
\section{The OAM of light}
Photons can carry orbital angular momentum, OAM \cite{ABP03}. We can define single photon states $\ket{\ell}$ which carry an orbital angular momentum of $\ell \hbar$. These states correspond to Laguerre-Gauss modes $LG_{p}^{\ell}$ which are orthonormal for different azimuthal integer $\ell$ indices. We assume a constant radial index $p=0$. OAM states are characterized by a spiral phase front with $\ell$ complete changes of phase from $0$ to $2\pi$ in the azimuthal coordinate. 
 
\section{OAM toolbox}
We can manipulate photons carrying OAM with standard linear optics equipment like phase shifters and beamsplitters with the usual results. Mirrors flip the phase front and change the sign of the winding number $\ell$. Additionally, we can specifically address the OAM degree of freedom with two new elements: Dove prisms and holograms.

A Dove prism rotated by an angle $\frac{\alpha}{2}$ rotates the wavefront, introducing a phase shift proportional to the OAM value of the photon state. We represent this element with the operator $D_{\alpha}$ so that $D_{\alpha}\ket{\ell}=e^{-i\alpha \ell}\ket{-\ell}$ \cite{GMT06}. The sign change in $\ell$ is due to reflection inside the prism.

We can also increase or decrease the OAM value of an $\ket{\ell}$ state by an arbitrary amount using a computer generated hologram \cite{ADA98}. Even though we use the term hologram, there are other optical elements, like spiral phase plates \cite{TRS96}, which produce the same result. We say we have a $+k$ hologram, with operator $H_{k}$, if the state evolution is $H_{k}\ket{\ell}=\ket{\ell+k}$. 

\section{Information encoding}
We consider networks with $D$ users transmitting photonic qubits. The data is encoded in individual photons with three relevant degrees of freedom: path, polarization and OAM. $\ket{\ell^p}_n$ denotes the state of a photon in path $n$ with polarization $p$ and winding number $\ell$. The users encode their data in the polarization degree of freedom. We consider user $i$ can generate arbitrary qubit states $\ket{\psi_i^{\ell}}=\alpha_i \ket{\ell^H}_i+\beta_i \ket{\ell^V}_i$, with $|\alpha_i|^2+|\beta_i|^2=1$. All the schemes will also be valid for entangled inputs in cases where many users share a state. 

We suppose all the blocks are insensitive to polarization. Our setups, however, can have small differences in behaviour for $\ket{\ell^H}_i$ and $\ket{\ell^V}_i$ states. In a practical implementation, it might be better to use alternative degrees of freedom to encode qubits. A possible substitute is time-bin encoding, the encoding used in practical quantum cryptography optical fibre networks \cite{BGT99}. In time-bin encoding, different qubit values are represented in separate time windows. The wavefunction of a time localized photon, or a part of it, can be delayed with respect to a reference to give two logical states.

\section{Orbital angular momentum beamsplitter}
The central element of all of our proposals is the OAM beamsplitter of Zou and Mathis \cite{ZM05}. An OAMBS forwards photons to different output ports according to their OAM. The OAMBS is a linear optics multiport with $D$ input ports and $D$ output ports. The $D$ paths can contain photons with any winding number $\ell$ from $0$ to $D-1$. Strictly, the OAMBS has $D\times D \times 2= 2D^2$ modes ($D$ paths, with $D$ OAM states and 2 polarizations), but we can study the evolution of a photon going through the OAMBS using the $D\times D$ scattering matrix that defines the $D$-path multiport. Except for reflections, the winding number $\ell$ of the photon does not change inside the passive, linear OAMBS and can be left as a parameter. Polarization is likewise preserved.

The OAMBS has two symmetric multiport gates, $S$, and a Dove prism stage, $D\!P$, that has at each port a Dove prism rotated by an angle $\frac{\pi n}{D}$ proportional to the port number $n$. The whole system has a scattering matrix $O\!A\!M\!B\!S=S D\!P S$. Matrix $S$ has elements
\begin{equation}
S_{n,m}=\frac{1}{\sqrt{D}}e^{i\frac{2\pi n m}{D}}
\label{Sel}
\end{equation}
in row $n$ and column $m$ and matrix $D\!P$ is a diagonal matrix with elements
\begin{equation}
D\!P_{n,n}=e^{-i\frac{2\pi \ell n}{D}}
\end{equation}
in the $n$-th row and column. 

\begin{figure}[ht]
\centering
\includegraphics[scale=0.75]{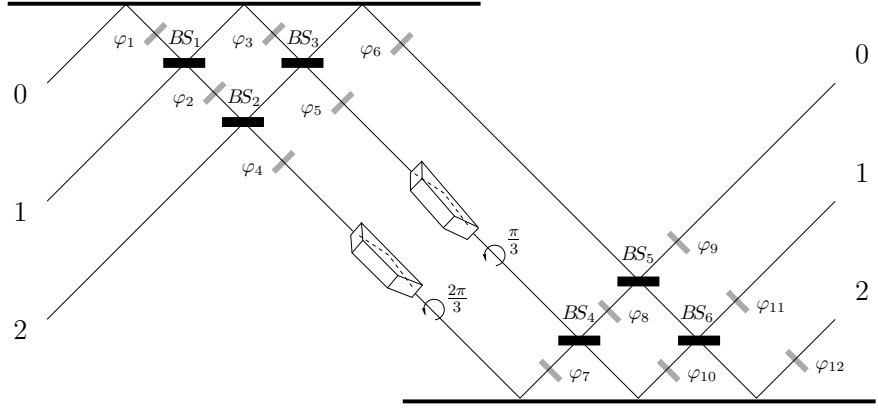}
\caption{Optical implementation of an OAMBS with 3 input/output paths. Two symmetric multiports implemented in a triangular configuration are connected through a Dove prism stage to separate the path of photons with different OAM. \label{OAMBS}} 
\end{figure}

Figure \ref{OAMBS} shows an OAMBS with three inputs and outputs. The symmetric multiports are built using beamsplitters, phase shifters and a mirror. Connecting the two symmetric multiports there are two Dove prisms which introduce different phase shifts for photons with a different winding number. All the elements inside the OAMBS are simple linear optics elements.

This kind of setup can be generalized to any number of inputs $D$. Symmetric multiports of any size can be synthesized with beamsplitters and phase shifters in a triangular setup \cite{RZB94}. We only need to add $D-1$ Dove prisms with the adequate rotation between two equal symmetric multiports to have an OAMBS. Inside the symmetric multiports the photons are reflected an odd number of times. The output photon will change its OAM state from $\ket{\ell}$ to $\ket{-\ell}$. The effect of the whole OAMBS acting on an input $\ket{\ell}_n$ state comes from applying the gates in order
\begin{eqnarray}
\label{ev1}
\ket{\ell}_n&\stackrel{S}{\rightarrow}&\frac{1}{\sqrt{D}}\sum_{m=0}^{D-1}e^{i\frac{2\pi n m}{D}}\ket{-\ell}_m\\
&\stackrel{D\!P}{\rightarrow}&\frac{1}{\sqrt{D}}\sum_{m=0}^{D-1}e^{i\frac{2\pi (n+\ell) m}{D}}\ket{\ell}_m \\
&\stackrel{S}{\rightarrow}&\frac{1}{D}\sum_{m=0}^{D-1}\sum_{k=0}^{D-1}e^{i\frac{2\pi m (n +\ell+k)}{D}}\ket{-\ell}_k=\ket{-\ell}_{\ominus \ell \ominus n}
\label{ev3}
\end{eqnarray}

The total probability amplitude for each $\ket{-\ell}_k$ state can be written as the geometric sum $\frac{1}{D}\sum_{m=0}^{D-1}\left(e^{i\frac{2\pi(n +\ell+k)}{D}}\right)^m$. Only the state $\ket{-\ell}_k$ with $k=rD-\ell-n$ has a probability amplitude different from 0. We use the shorthand $\ket{-\ell}_{\ominus\ell \ominus n}$, where $\ominus$ is modulo $D$ subtraction. In the rest of the paper, whenever modular addition, $\oplus$, or subtraction, $\ominus$, appear, they are assumed to be modulo $D$.  

Photons will also travel through the OAMBS in the opposite direction. In the previous discussion we have described the left-to-right or direct evolution with operator $O\!A\!M\!B\!S$. We will also send photons from the ``output'' ports. We call the operator in the right-to-left or reverse direction $S\!B\!M\!AO$. 

We can relate $O\!A\!M\!B\!S$ and $S\!B\!M\!AO$ using basic linear algebra properties. Matrix $S$ is the concatenation of beamsplitter, $B\!S$, and phase shifter, $P\!S$, stages so that $S=P\!S_N B\!S_N\cdots B\!S_1 P\!S_1$. All the stages have symmetric scattering matrices ($B\!S_i^T=B\!S_i$ and $P\!S_i^T=P\!S_i$). $S^T=(P\!S_N B\!S_N\cdots B\!S_1 P\!S_1)^T=P\!S_1 B\!S_1\cdots B\!S_N P\!S_N$ is the scattering matrix that corresponds to applying the stages in reverse order (from right to left). As $S$ is symmetric, see Equation (\ref{Sel}), the $S$ operators are the same in the forward and backward directions. 

The Dove prism acts only on the OAM. In this case, the backwards operation is different. The winding number is defined from the point of view of the direction of propagation. In a right to left journey, photons see a rotation angle $\frac{-\alpha}{2}$ instead of $\frac{\alpha}{2}$. For the same $\ell$, we find diagonal matrix elements $e^{i\frac{2\pi \ell n}{D}}$ and an operator $D\!P^\dag$. We can repeat the calculations in equations (\ref{ev1})-(\ref{ev3}) to see that evolution through $S\!B\!M\!AO=S D\!P^{\dag} S$ is 
\begin{eqnarray}
\ket{\ell}_n&\stackrel{S}{\rightarrow}&\frac{1}{\sqrt{D}}\sum_{m=0}^{D-1}e^{i\frac{2\pi n m}{D}}\ket{-\ell}_m\\
&\stackrel{D\!P^\dag}{\rightarrow}&\frac{1}{\sqrt{D}}\sum_{m=0}^{D-1}e^{i\frac{2\pi (n-\ell) m}{D}}\ket{\ell}_m \\
&\stackrel{S}{\rightarrow}& \frac{1}{D}\sum_{m=0}^{D-1}\sum_{k=0}^{D-1}e^{i\frac{2\pi m (n -\ell+k)}{D}}\ket{-\ell}_k=\ket{-\ell}_{\ell \ominus n}\!.
\end{eqnarray}
Now, only the state $\ket{\ell}_k$ with $k=rD+\ell-n$ has a probability amplitude different from 0.

We can check that an SBMAO is the inverse of the OAMBS, as
\begin{equation}
\ket{\ell}_n\stackrel{O\!A\!M\!B\!S}{\rightarrow}\ket{-\ell}_{\ominus \ell \ominus n}\stackrel{S\!B\!M\!AO}{\rightarrow} \ket{\ell}_n.
\end{equation} 

\section{Compact OAM multiplexing}
\begin{figure}[ht]
\centering
\includegraphics[scale=0.5]{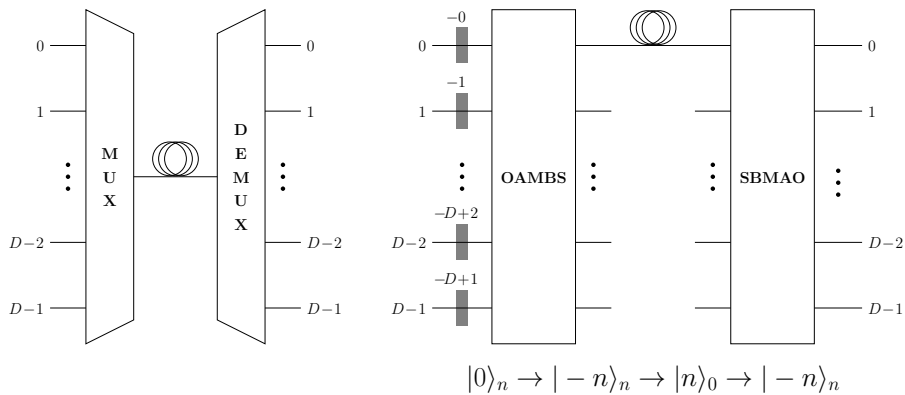}
\caption{\emph{Left:} A multiplexer (MUX) combine many channels into one line to make transmission easier. At the receiver, a demultiplexer (DEMUX) redistributes the data to the appropriate users. \emph{Right}: A combination of holograms with an OAMBS can be used as a MUX. An SBMAO at the receiving end works as a DEMUX.\label{MUX}} 
\end{figure}

Our first network application is a compact OAM merger which takes photons in different paths and combines them in a single spatial port. The merger follows the general scheme of Figure \ref{MUX}. This system is different from our previous OAM multiplexer, which takes the state from $n$ photonic qubits and transfers it to the OAM of a single photon. That kind of multiplexing requires advanced photonic quantum gates beyond current technology. We have also proposed a limited qubit merger which can take together $n$ photons and send them through the same path, but is limited to an exponentially growing OAM encoding \cite{GC08}. 

The OAM MUX we describe here can take up to $D$ photons into a single path, but is more efficient and experimentally feasible for a higher number of channels. We use all the OAM values from 0 to $D$. 

Figure \ref{MUX} shows a general multiplexer and our OAM implementation with OAMBSs. A $D\times 1$ multiplexer, MUX, takes $D$ different inputs into a single path and, at the other side of the channel, a $1\times D$ demultiplexer, DEMUX, separates the photons into $D$ output paths. The $D\times 1$ MUX is built from an OAMBS with a hologram before each input port. We imagine $D$ users who are assigned each an input port of index $n$ from $0$ to $D-1$. We then place a $-n$ hologram before the $n$-th port of the OAMBS. Input user states $\ket{0}_n$ become $\ket{-n}_n$, which, after the OAMBS, become $\ket{n}_0$. All the input photons are in the path corresponding to the $0$ output port and have an identifying OAM value which allows to separate them later. If the photons carry polarization encoded quantum data, the transmitted multiplexed state can be written as $\otimes_{n=0}^{D-1}\left(\alpha_n\ket{n^H}_0+\beta_n\ket{n^V}_0\right)$. The DEMUX is simply a SBMAO (the input OAMBS with the elements in the reverse order). The SBMAO takes states $\ket{n}_0$ into $\ket{-n}_n$. 

As the data is encoded in the polarization degree of freedom, we don't really need to undo the OAM transformation. Anyway, a final stage with $+n$ holograms can be added at output ports $n$ for symmetry or if the OAM state of the photons can be relevant later.

\section{Self-routing networks}
Self-routing networks are systems in which the messages, as sent from the users, carry all the information needed to deliver them to their desired destination. Intermediate nodes perform little additional processing and the network is simple. The best known classical example are Benes networks \cite{NS81}.

\subsection{Simple OAMBS self-routing}
We can use the OAM degree of freedom of the user photons to encode the destination node so that a simple optical network can forward them correctly. There are two possible configurations. The first is a direct application of the OAMBS. If we have two sets of users, a group of $D$ ``left'' users $L_0, L_1, \ldots, L_{D-1}$ and a group of $D$ ``right'' users $R_0, R_1, \ldots, R_{D-1}$, any user from the left group can send data to any right user of choice. Conversely, any right user can reach any left user. If user $L_n$ prepares a state $\ket{\ell}_n$, the output state $\ket{-\ell}_{\ominus \ell\ominus n}$ is received by user $R_{\ominus \ell\ominus n}$. A user $L_n$ who wants to send a qubit to user $R_m$ can always prepare a photon with winding number $\ell=\ominus m\ominus n$. Users need to have variable holograms, which can be implemented, for instance, with spatial light modulators, SLMs, controlled by a computer with the pre-stored patterns of each necessary $\pm \ell$ value \cite{SGJ07}. The network also works in reverse to send any message from $R_n$ to $L_m$. In this case, the photon sees the $S\!B\!M\!AO$ transformation. $R_n$ must choose a winding number $\ell=m\oplus n$ that directs input state $\ket{m\oplus n}_n$ into an output state $\ket{-(m\oplus n)}_m$ in path $m$ where $L_m$ is. 

\subsection{OAM star network}
\begin{figure}[ht]
\centering
\includegraphics[scale=0.75]{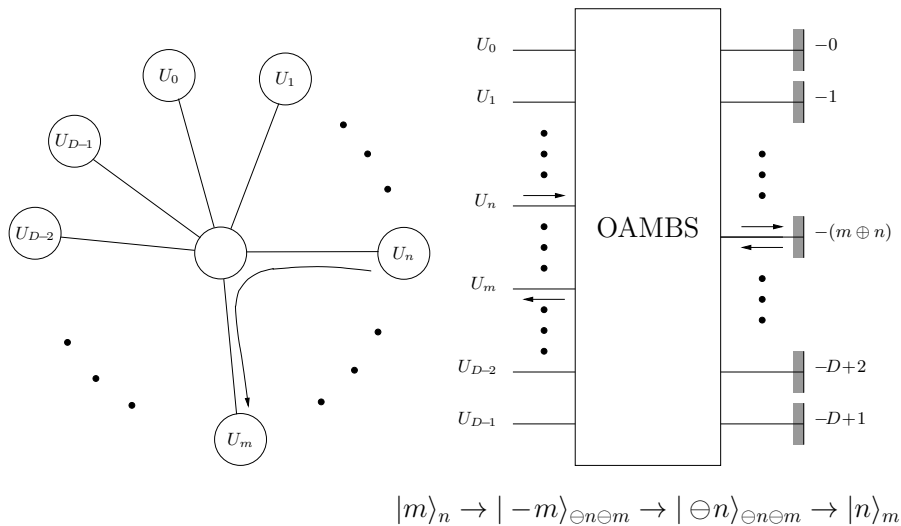}
\caption{Star self-routing network with OAM routing. An OAMBS with reflective holograms at the output works as a redistributing node. Users can reach any destination by sending a photon with the appropriate OAM value to the OAMBS.\label{star}} 
\end{figure}

We can improve the previous design to build a self-routing network where all the users can send qubits to any other user. Figure \ref{star} shows the diagram representing these star networks with a central distributing node. 

We can build this network by using a modified version of the OAMBS setup. We make use of reflective holograms, such as those created with reflective SLMs which combine a mirror and a hologram. The effect of a reflective hologram $R\!H_k$ on OAM state $\ket{\ell}_n$ is $H\!L_k\ket{\ell}_n=\ket{-\ell-k}_n$. The star network is created for $D$ users at the input of an OAMBS. We place reflective holograms $H\!L_{-n}$ at the $n$-th output port of the OAMBS. 

Imagine user $U_n$ wants to send a qubit to user $U_m$. $U_n$ can prepare a state $\ket{\ell}_n$, which, after the OAMBS, becomes $\ket{-\ell}_{\ominus \ell\ominus n}$. The state is reflected from the hologram with $k= \ell\oplus n$ and state $\ket{\ominus n}_{\ominus \ell\ominus n}$ transverses the OAMBS in the reverse direction. After the $S\!B\!M\!AO$ evolution, the state is $\ket{n}_{\ominus n\oplus \ell\oplus n}=\ket{n}_{\ell}$. User $U_n$ only needs to give its qubit a winding number $\ell=m$ to make it reach user $U_m$.

\section{Summary and further possibilities}
We have presented two different applications of orbital angular momentum to quantum networking. First we have given an OAM multiplexer which can combine up to $D$ photonic qubits in a single output port and the corresponding OAM demultiplexer that expands the qubits back to $D$ destination ports. Then, we have proposed two different self-routing networks in which the OAM degree of freedom of the transmitted photons guarantees they are delivered to the desired destination. All the proposed systems can be built with simple optical elements which have been used before with good results in experiments on the OAM of single photons \cite{LPB02,LJR09}. 

There are some practical difficulties that limit our proposals. The first is that OAM modes do not couple well to optical fibre. OAM encoding is mostly restricted to free-space communication networks. Second, there is a maximum winding number $\ell$ beyond which the generation and manipulation of photons in the $\ket{\ell}$ state becomes too technically demanding. Nevertheless, OAM values up to a few tens have been successfully employed in practical communication scenarios \cite{GCP04}. The multiplexing and routing techniques we have proposed could be used in a small or medium open network. 

We can picture a metropolitan quantum cryptography network with a central node on the top of a visible, tall building. If the central node contains the OAMBS and the reflective holograms of the star self-routing network, any pair of users can establish communication channels between them without dedicated links for each possible connection. Visibility between users is no longer required. They only need to see the central node. For $D$ users, the number of nodes goes down from $D \choose 2$ to $D$ (plus the additional optics in the central node with the star router). Similar networks could be deployed, for instance, for satellite quantum cryptography \cite{RTG02,AJP03}.

The networks we have proposed can also be used to send classical information. In classical information networks, OAM gives some advantages, like improved security against eavesdropping the OAM value \cite{GCP04}. Quantum OAM networks, however, open many intriguing possibilities, like new network protocols with a superposition of destinations. 

A simple interesting application is entanglement distribution \cite{CZK97,HB02}. Imagine a polarized entangled Bell pair $\frac{1}{\sqrt{2}}\left(\ket{0^H}_x\ket{0^V}_y+\ket{0^V}_x\ket{0^H}_y\right)$ with two paths which can be, for instance, the paths of the photon pairs emerging from parametric down conversion in a BBO crystal \cite{SA88}. The crystal can be connected to the two input ports corresponding to users $U_x$ and $U_y$. Before entering our OAM star network, we can place $+n$ and $+m$ holograms to reach destination users $U_n$ and $U_m$. After the journey back and forth the OAMBS, the end users share the entangled state $\frac{1}{\sqrt{2}}\left(\ket{x^H}_n\ket{y^V}_m+\ket{x^V}_n\ket{y^H}_m\right)$. With two SLMs in the source of the Bell pairs we can change the destination nodes. In any case, due care must be taken to preserve entanglement. Too long a network journey is still problematic without quantum repeaters \cite{BDC98}.

These examples show the potential of having a network with automated routing at the quantum level. We hope our OAM quantum network elements will clear the way for new experimental quantum computer networks in which the network tasks can also take advantage of different quantum effects. 

\section{Acknowledgements}
 This research has been funded by Junta de Castilla y Le\'on project VA342B11-2 and MICINN TEC2010-21303-C04-04.

\newcommand{\noopsort}[1]{} \newcommand{\printfirst}[2]{#1}
  \newcommand{\singleletter}[1]{#1} \newcommand{\switchargs}[2]{#2#1}

\end{document}